\documentclass[submission, Proceedings]{SciPost}

\hypersetup{
    colorlinks,
    linkcolor={red!50!black},
    citecolor={blue!50!black},
    urlcolor={blue!80!black}
}

\usepackage[bitstream-charter]{mathdesign}
\usepackage{bm}

\newcommand{\bea}{\begin{eqnarray}}
\newcommand{\eea}{\end{eqnarray}}
\newcommand{\be}{\begin{equation}}
\newcommand{\ee}{\end{equation}}

\DeclareSymbolFont{usualmathcal}{OMS}{cmsy}{m}{n}
\DeclareSymbolFontAlphabet{\mathcal}{usualmathcal}

\begin{document}

\begin{center}{\Large \textbf{
$\Lambda$ polarizing fragmentation function from Belle $e^+e^-$ data
}}\end{center}

\begin{center}
Umberto D'Alesio\textsuperscript{1,2$\star$},
Francesco Murgia\textsuperscript{2} and
Marco Zaccheddu\textsuperscript{1,2}
\end{center}

\begin{center}
{\bf 1} Dipartimento di Fisica, Universit\`a di Cagliari,  Cittadella Universitaria di Monserrato, I-09042 Monserrato (CA), Italy
\\
{\bf 2}  INFN, Sezione di Cagliari,  Cittadella Universitaria di Monserrato, I-09042 Monserrato (CA), Italy
\\
* umberto.dalesio@ca.infn.it
\end{center}

\begin{center}
\today
\end{center}


\definecolor{palegray}{gray}{0.95}
\begin{center}
\colorbox{palegray}{
  \begin{tabular}{rr}
  \begin{minipage}{0.1\textwidth}
    \includegraphics[width=22mm]{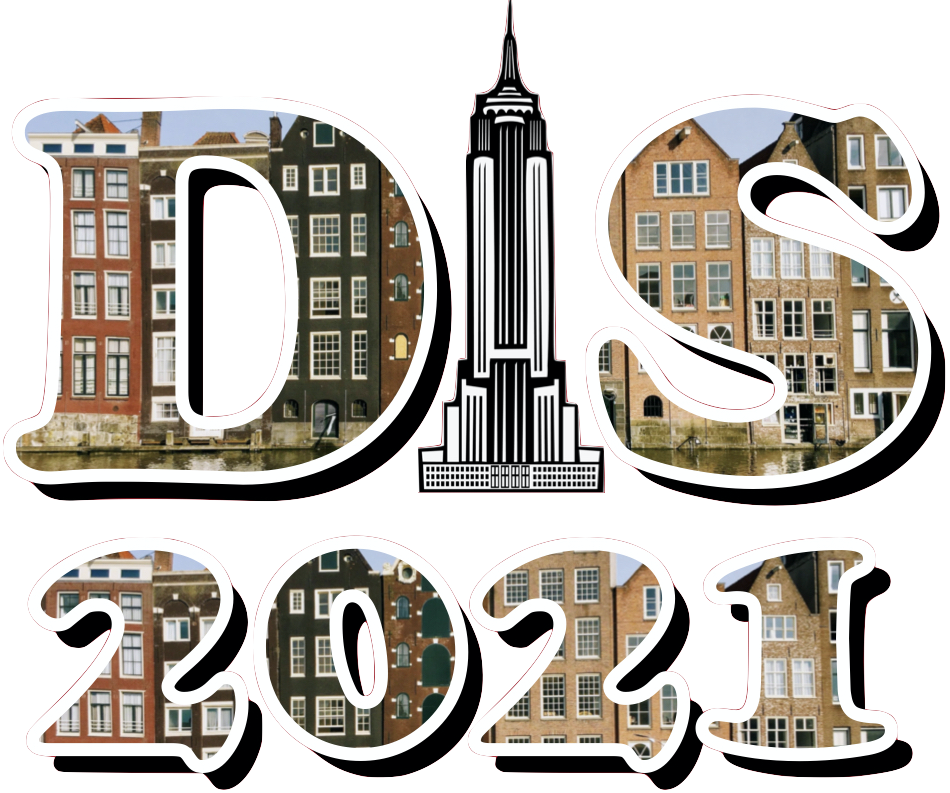}
  \end{minipage}
  &
  \begin{minipage}{0.75\textwidth}
    \begin{center}
    {\it Proceedings for the XXVIII International Workshop\\ on Deep-Inelastic Scattering and
Related Subjects,}\\
    {\it Stony Brook University, New York, USA, 12-16 April 2021} \\
    \doi{10.21468/SciPostPhysProc.?}\\
    \end{center}
  \end{minipage}
\end{tabular}
}
\end{center}

\section*{Abstract}
{\bf
Experimental data from Belle Collaboration for the transverse polarization of $\Lambda$'s measured in $e^+ e^-$ annihilation processes are used to extract the polarizing fragmentation function (FF) within a TMD approach.
We consider both associated and inclusive $\Lambda$ production, showing a quite consistent scenario. Good separation in flavor is obtained, leading to four independent FFs. Predictions for SIDIS processes at the EIC, crucial for understanding their universality and evolution properties, are also presented.
}

\section{Introduction}
\label{sec:intro}

Transverse momentum dependent parton distribution and fragmentation functions (TMDs) play a crucial role in our understanding of the internal structure of the nucleon as well as the parton fragmentation mechanism into hadrons. In this contribution we focus on one TMD that, despite its relevance, has been still poorly explored: the so-called polarizing fragmentation function (pFF), giving the distribution of a transversely polarized spin-1/2 hadron in the fragmentation of an unpolarized quark. It is T-odd, but chiral even, allowing to access it without any unknown, chiral-odd, counterpart.

It was studied phenomenologically long time ago in the context of the puzzling transverse polarization data for the inclusive production of $\Lambda$'s in unpolarized hadron-hadron collisions~\cite{Anselmino:2000vs}, but the lack of additional experimental information prevented any further theory development. The situation has significantly changed with the release of new data from the Belle Collaboration at KEK~\cite{Guan:2018ckx} on transverse $\Lambda/\bar\Lambda$ hyperon polarization in $e^+e^-$ processes. Indeed, for processes like $e^+e^-\to \Lambda\, h + X$, in contrast to the inclusive process $pp\to \Lambda + X$, a TMD factorization approach has been formally proven~\cite{Collins:2011zzd,GarciaEchevarria:2011rb}.

We report here the findings of Ref.~\cite{DAlesio:2020wjq}, showing how Belle data allow for the first ever extraction of the pFF in a clear way.
We will first consider the associated production ($\Lambda\,h$) data set alone, for which TMD factorization holds. We then include, within a simplified TMD approach, the $\Lambda$-jet case, by performing a separate full-data fit. Even if this configuration presents some caveats both experimentally and theoretically~\cite{Kang:2020yqw,Boglione:2020auc}, it represents a clear source of information on the explicit $p_\perp$ dependence of the TMD pFF.

\section{Formalism}
We consider the processes $e^+e^-\to h_1 h_2 + X$ where $h_1$ is a spin-1/2 hadron and the second (light and unpolarized) hadron, $h_2$, is produced almost back-to-back with respect to $h_1$. The complete calculation, within the helicity formalism, of all accessible spin observables will be presented elsewhere. Here we discuss in detail the case of the transverse hyperon polarization, defined generically as
\begin{equation}
{\cal P}_T = \frac{d\sigma^{\uparrow} - d\sigma^{\downarrow} }{d\sigma^{\uparrow} + d\sigma^{\downarrow}} = \frac{d\sigma^{\uparrow} - d\sigma^{\downarrow} }{d\sigma^{\rm unp}} ,
\label{polt}
\end{equation}
where $d\sigma^{\uparrow(\downarrow)}$ is the differential cross section for the production of a hadron transversely polarized along the up(down) direction with respect to the production plane and $d\sigma^{\rm unp}$ is the unpolarized cross section.

As we will show below, this observable is directly sensitive to the polarizing FF, $\Delta D_{h_1^\uparrow\!/q}$, also denoted as $D_{1T}^{\perp q}$~\cite{Bacchetta:2004jz}. More precisely, for a hadron with a lightcone momentum fraction $z$, intrinsic transverse momentum $\bm{p}_\perp$ and polarization vector $\hat{\bm{P}}\equiv \uparrow$, coming from the fragmentation of a quark with momentum $\bm{p}_q$, the pFF is defined as
\bea
\label{polFF}
\Delta \hat D_{h^\uparrow\!/q}(z,\bm{p}_{\perp}) \equiv \hat D_{h^\uparrow\!/q}(z,\bm{p}_\perp) - \hat D_{h^\downarrow\!/q}(z,\bm{p}_\perp) = \Delta D_{h^\uparrow\!/q}(z,p_{\perp}) \,\hat{\bm{P}} \cdot (\hat{\bm{p}}_q \times \hat{\bm{p}}_\perp) \,.
\eea

For associated hadron production the following configuration is adopted: the produced unpolarized hadron, $h_2$, defines the opposite $z$ direction 
and the $\widehat{xz}$ plane is determined by the lepton and the $h_2$ directions. 
A second plane is determined by $\hat{\bm z}$ and the direction of the spin-1/2 hadron, $h_1$, with three-momentum $\bm{P}_{h_1}= (P_{1T} \cos\phi_1,P_{1T}\sin\phi_1,P_{1L})$.
This is the \emph{hadron frame} configuration and the transverse polarization is measured along $\hat{\bm{n}} \propto (-\bm{P}_{h_2}\times \bm{P}_{h_1})$.

The transverse polarization of $h_1$, as defined in its helicity frame, has to be properly projected along $\hat{\bm{n}}$.
Moreover, two independent modulations appear: one in term of the convolution of the pFF with the unpolarized TMD-FF for $h_2$, and another one involving the Collins FF for $h_2$, that, however, vanishes upon integration over $\bm{P}_{1T}$.
The final expression for the polarization along $\hat{\bm{n}}$, so integrated, and adopting a Gaussian Ansatz for the TMD-FFs reads
\bea
{\cal P}_n(z_1,z_2) & = & \sqrt{\frac{e\pi}{2}}\frac{1}{M_{\rm pol}} \frac{\langle p_\perp^2\rangle_{\rm pol}^2}{\langle p_{\perp 1}^2\rangle}\,\frac{z_2}{\big\{[z_1(1-m_{h_1}^2/(z_1^2s))]^2\langle p_{\perp 2}^2\rangle +z_2^2 \langle p_\perp^2\rangle_{\rm pol}\big\}^{1/2}}\nonumber\\
&\times & \frac{\sum_{q} e^2_q\,\Delta D_{h_1^\uparrow/q}(z_1)D_{h_2/\bar q}(z_2)}{ \sum_{q}  e^2_q\,
 D_{h_1/q}(z_1)D_{h_2/\bar q}(z_2)}\,,
\label{PolTh}
\eea
where we have used the following parametrization for the TMD-FFs
\bea
D_{h/q}(z,p_{\perp}) & = & D_{h/q}(z)\, \frac{e^{-p_{\perp}^2/\langle p_{\perp}^2\rangle}}{\pi \langle p_{\perp}^2\rangle} \,, \\
\Delta D_{h^\uparrow\!/q}(z,p_{\perp}) &= & \Delta D_{h^\uparrow\!/q}(z)  \frac{\sqrt{2e} \,p_{\perp}}{M_{\rm pol}} \frac{e^{-p_{\perp}^2/\langle p_{\perp}^2\rangle_{\rm pol}}}{\pi \langle p_{\perp}^2\rangle},
\label{Gaus}
\eea
with $\langle p_\perp^2\rangle_{\rm pol} = \frac{M_{\rm pol}^2}{M_{\rm pol}^2 + \langle p_{\perp}^2\rangle} \,\langle p_{\perp}^2\rangle$.
By imposing $|\Delta D(z)|\le D(z)$ the positivity bound for the pFF, Eq.~(\ref{polFF}), is automatically fulfilled.

As mentioned above, we consider also the inclusive $\Lambda$ production case (within a jet) in a simplified TMD approach. In this case the polarization is measured orthogonally to the \emph{thrust plane}, containing the jet and the spin-1/2 hadron momentum, $\bm{P}_{h_1}$.
Skipping all details, the transverse polarization is then given as
\bea
{\cal P}_T(z_1, p_{\perp 1}) &  = & \frac{\sum_{q} e^2_q\,
\Delta D_{h_1^\uparrow/q}(z_1,p_{\perp 1})}{ \sum_{q}  e^2_q\,
D_{h_1/q}(z_1,p_{\perp 1})}\,.
\label{PolTjet}
\eea
For later use we give the first $p_\perp$-moment of the pFF:
\be
\Delta D_{h^\uparrow\!/q}^{(1)}(z)  =  \int\! d^2 \bm{p}_{\perp} \frac{p_{\perp}}{2 z m_h} \Delta D_{h^\uparrow\!/q}(z,p_{\perp}) =
D_{1T}^{\perp (1)}(z)
= \sqrt{\frac{e}{2}}\frac{1}{z m_h} \frac{1}{M_{\rm pol}}\frac{\langle p^2_{\perp} \rangle^{2}_{\rm pol}}{\langle p^2_{\perp} \rangle}  \Delta D_{h^\uparrow\!/q }(z)\,,
\label{1mom}
\ee
where the last expression is obtained by using Eq.~(\ref{Gaus}).

\section{Results}
\label{results}

Two data sets are available: one for the associated production of $\Lambda$ with light hadrons ($\pi$ and $K$), as a function of the energy fractions $z_\Lambda$ and $z_\pi(z_K)$  (128 data points) and one for the inclusive production as a function of $p_{\perp}$ (the $\Lambda$ transverse momentum w.r.t.~the thrust axis), for different energy fractions, $z_\Lambda$ (32 data points).
%

The $z$-dependent part of the pFF is parameterized as follows
\be
\Delta D_{\Lambda^\uparrow\!/q}(z) = N_q z^{a_q} (1-z)^{b_q} \frac{(a_q+b_q)^{(a_q+b_q)}}{a_q^{a_q}b_q^{b_q}} D_{\Lambda/q}(z)\,,
\label{Deltaz}
\ee
where $|N_q|\le 1$ and $q=u,d,s, \rm{sea}$. This guarantees that $|\Delta D(z)|\le D(z)$.

For the light hadron unpolarized FFs we adopt the DSS07 set~\cite{deFlorian:2007aj}, and the AKK08 set~\cite{Albino:2008fy} for $\Lambda$'s. The unpolarized Gaussian widths are taken as $\langle p^2_{\perp} \rangle = 0.2$ GeV$^2$~\cite{Anselmino:2005nn}, for all hadrons. Evolution effects do not play any role in this extraction, being performed at a fixed energy scale.
Since all $\Lambda$ FF sets are given for $\Lambda+\bar\Lambda$, including the AKK08 set, we separate the two contributions assuming
\be
D_{\bar\Lambda/q} = D_{\Lambda/\bar q} = (1-z) \, D_{\Lambda/q}  \,.
\ee
Other choices have a very little impact on the fit.

As already stated, we start performing a fit of the associated production data alone (for a similar analysis see Ref.~\cite{Callos:2020qtu}), including in a second phase also the inclusive data set. To improve the quality of the fit, we have imposed the following cuts: $z_{\pi,K} \le 0.5$ for the associated production and $z_\Lambda \le 0.5$ for the $\Lambda$-jet data set, leaving us with 96+24 = 120 data points.

Concerning the $z$-dependent part, Eq.~(\ref{Deltaz}), we have tried many different combinations of parameter sets, leading to the following best choice:
\be
N_u, \; N_d, \; N_s, \; N_{\rm sea}, \; a_s, \; b_u, \; b_{\rm sea}\,,
\ee
with all other $a$ and $b$ parameters set to zero. This, together with $\langle p^2_{\perp} \rangle_{\rm pol}$ (Eq.~(\ref{Gaus})), implies  8 free parameters.

\begin{table}[t!]
\centering
\caption{
Best values of the 8 free parameters fixing the pFF (Eqs.~(\ref{Gaus}), (\ref{Deltaz})) for $u,d,s$ and sea quarks, as obtained by fitting the full set of Belle data~\cite{Guan:2018ckx}, with their statistical errors (see text).
\label{fitpar}}
\vskip 6pt
\renewcommand{\tabcolsep}{0.4pc} 
\renewcommand{\arraystretch}{1.} 
\begin{tabular}{@{}ll}
 \hline
 $N_{u} = 0.47^{+0.32}_{-0.20}$ & $N_{d} \;\;=  -0.32^{+0.13}_{-0.13}$ \\
 $N_{s} = -0.57^{+0.29}_{-0.43}$ & $N_{\rm sea} =  -0.27^{+0.12}_{-0.20}$\\
 $a_s \;= 2.30^{+1.08}_{-0.91}$ &\\
 $b_u \;= 3.50^{+2.33}_{-1.82}$ & $b_{\rm sea}  \;= 2.60^{+2.60}_{-1.74}$ \\
 \hline
  $\langle p^2_{\perp} \rangle_{\rm pol} = 0.10^{+0.02}_{-0.02}$ GeV$^2$ & \\
 \hline
\end{tabular}
\end{table}

\begin{figure}[!t]
\centering
\includegraphics[trim =  0 50 0 140,clip,width=6.5cm]{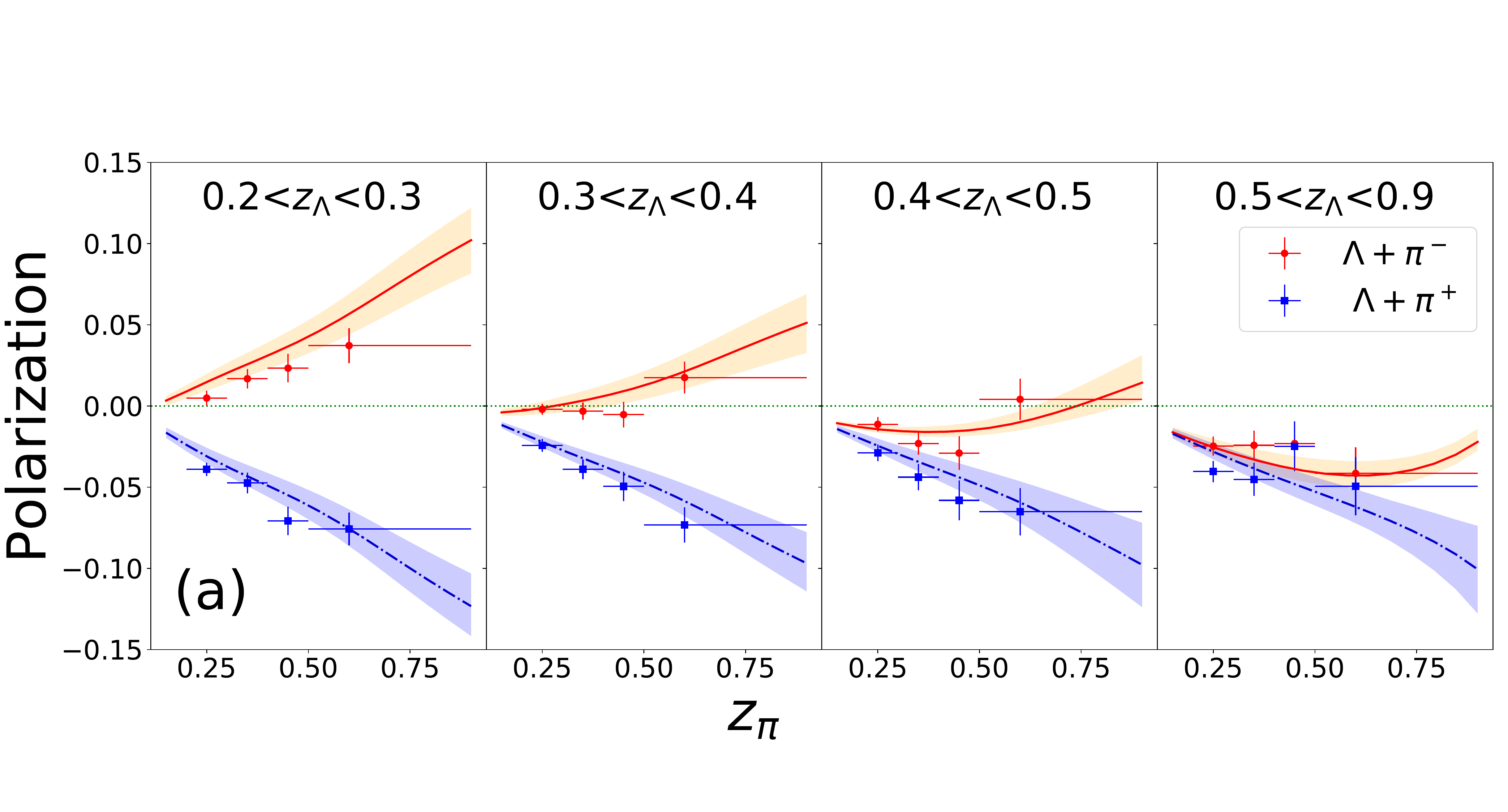}
\includegraphics[trim =  0 50 0 140,clip,width=6.5cm]{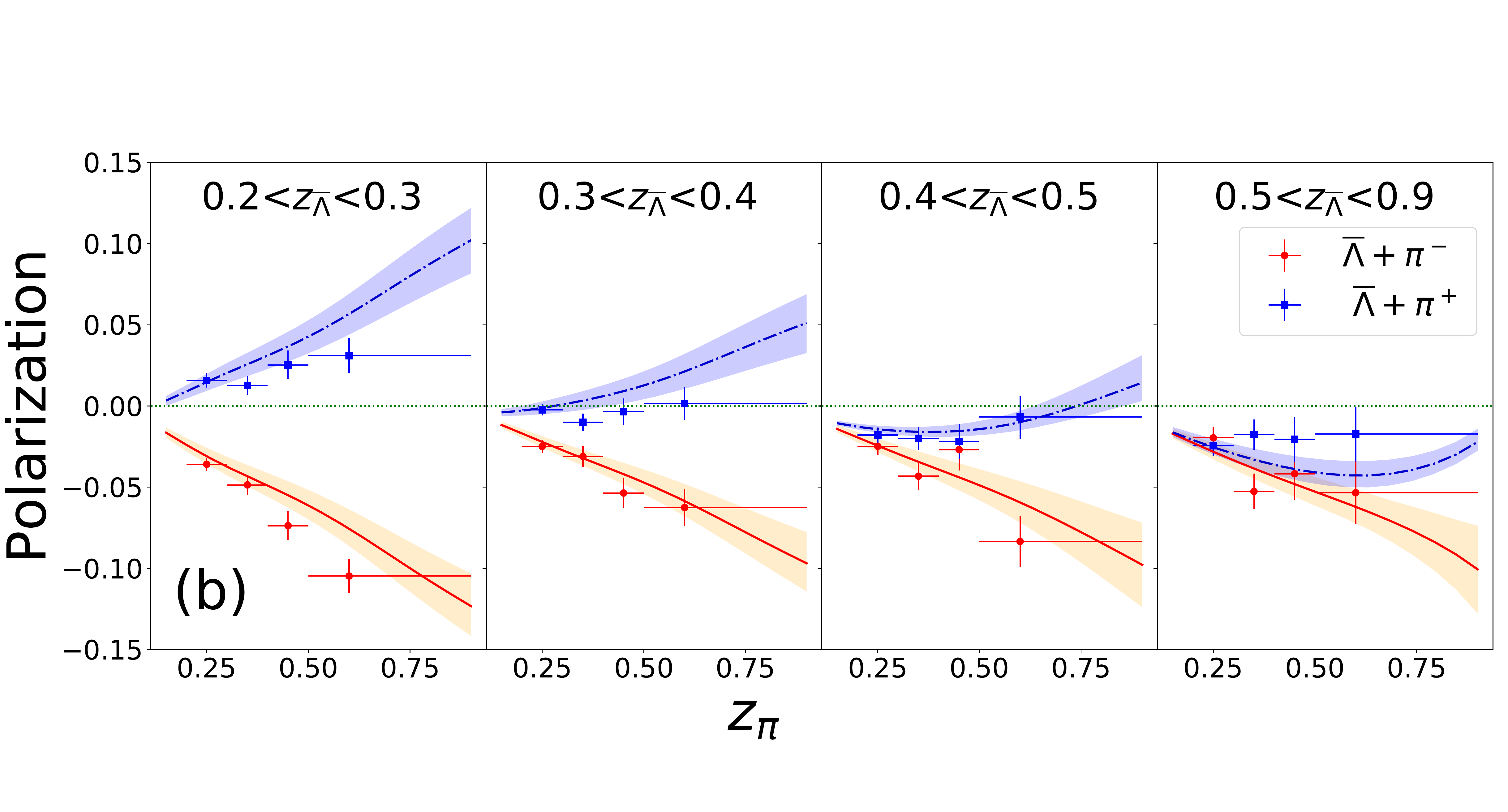}
\includegraphics[trim =  0 50 0 140,clip,width=6.5cm]{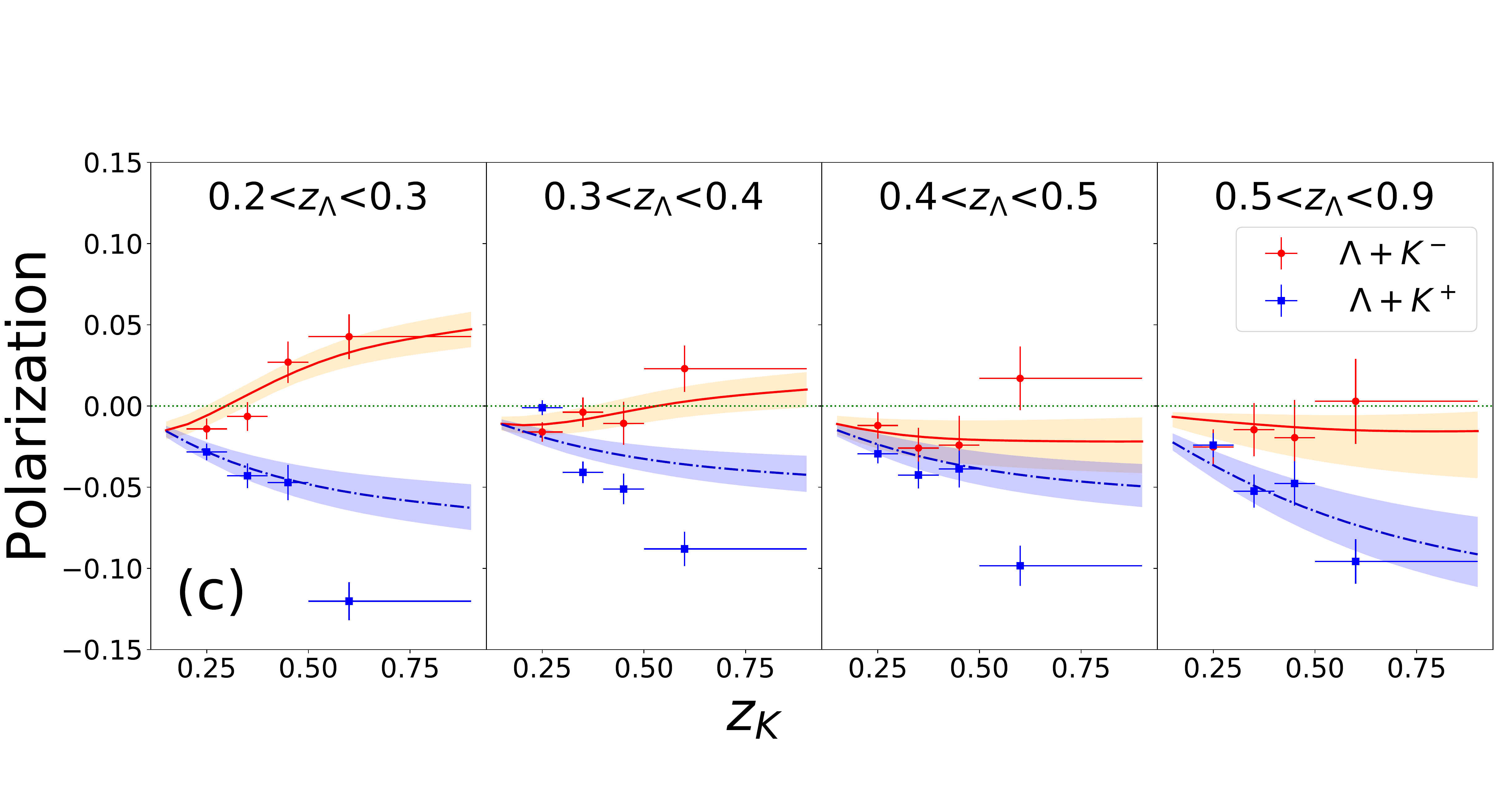}
\includegraphics[trim =  0 50 0 140,clip,width=6.5cm]{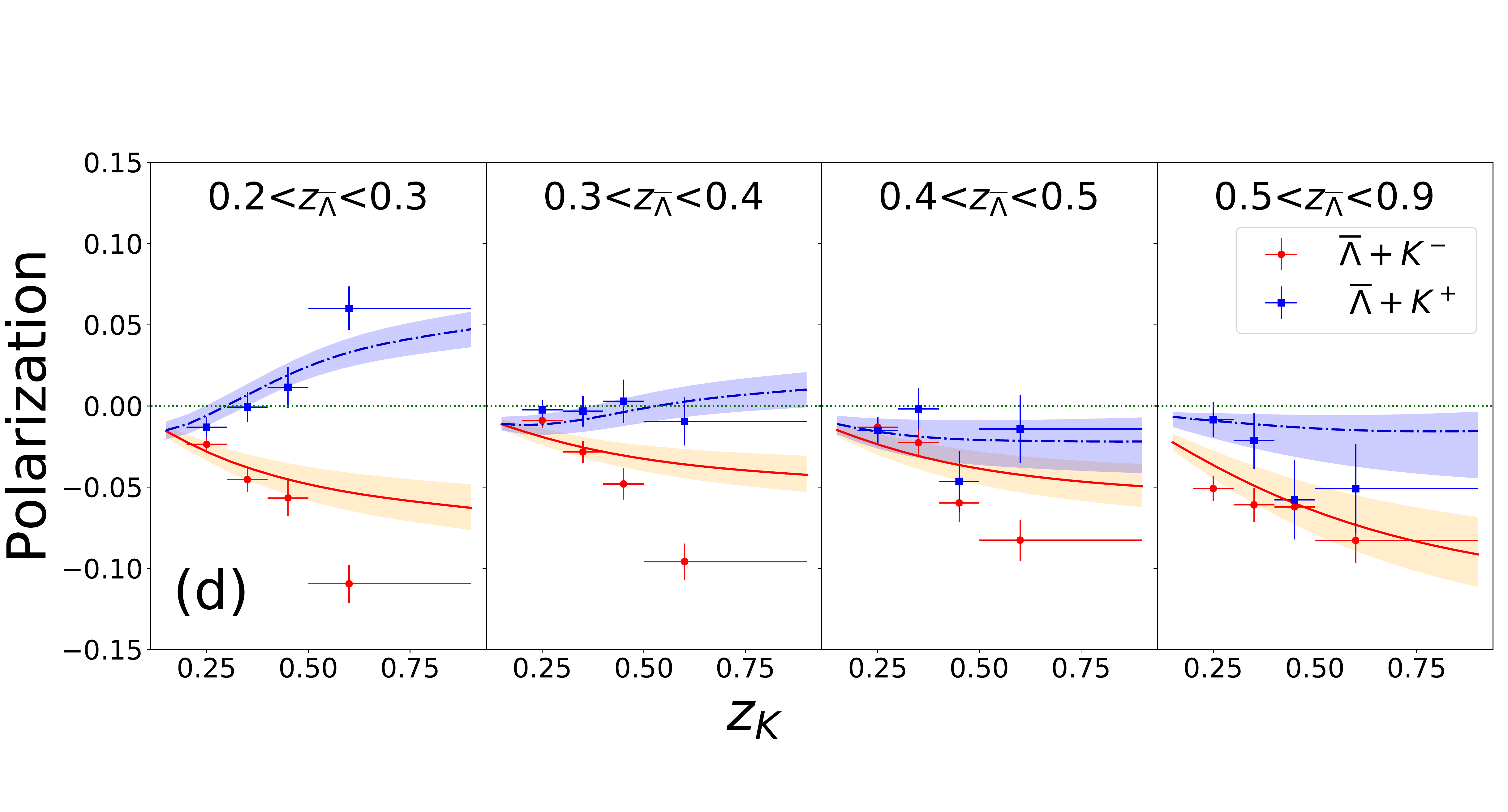}
\caption{Best-fit estimates, based on the full-data set, of the transverse polarization for $\Lambda$ and $\bar\Lambda$ production in $e^+e^-\to \Lambda(\bar\Lambda) h + X$, vs.~$z_{h}$ (of the associated hadron) for different $z_\Lambda$ bins. Data are from Belle~\cite{Guan:2018ckx}. The statistical uncertainty bands, at 2$\sigma$ level, are also shown. Data for $z_{\pi,K}>0.5$ are not included in the fit.}
\label{fig:Lh}
\end{figure}

\begin{figure}[!b]
\centering
\includegraphics[trim =  0 50 0 140,clip,width=6.5cm]{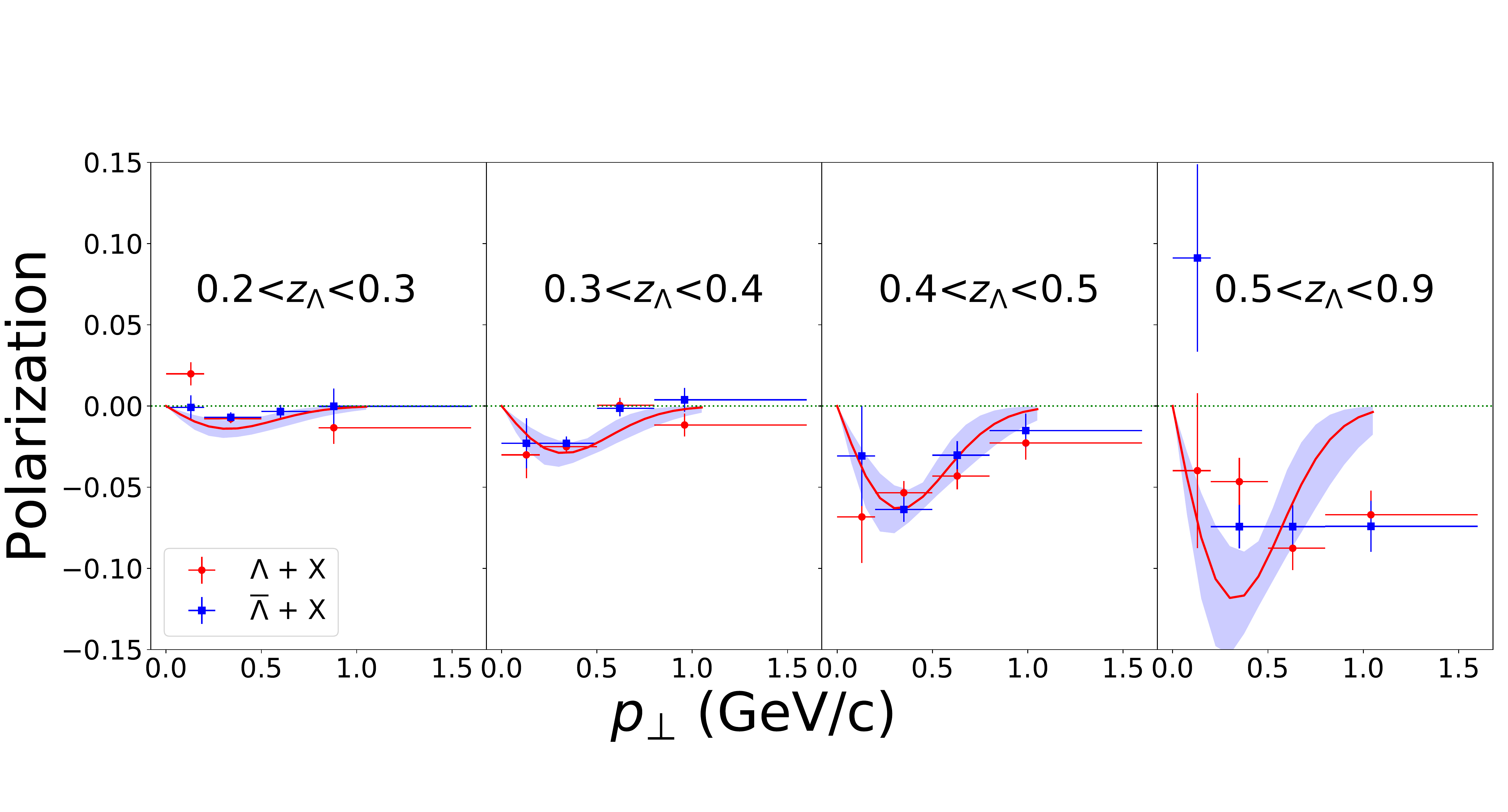}
\caption{Best-fit estimates of the transverse polarization for inclusive $\Lambda$/$\bar\Lambda$ production in $e^+e^-\to \Lambda(\rm{jet}) + X$ vs.~$p_\perp$ for different $z_\Lambda$ bins, compared against Belle data~\cite{Guan:2018ckx}. The statistical uncertainty bands, at 2$\sigma$ level, are also shown. Curves for $\Lambda$ and $\bar\Lambda$ coincide and data in the rightmost panel are not included in the fit.}
\label{fig:Lj}
\end{figure}

In Table~\ref{fitpar} we collect the values of the best-fit parameters for the full-data analysis. The corresponding estimates, compared against Belle data~\cite{Guan:2018ckx}, are shown in Figs.~\ref{fig:Lh} and \ref{fig:Lj}, respectively for the associated  and inclusive $\Lambda$ production. The fit of the associated production data leads to a $\chi^2_{\rm dof} = 1.26$ with $\chi^2_{\rm points}= 0.8, 1.5$ for pion and kaon data subsets, while the full-data fit gives a $\chi^2_{\rm dof} =1.94$,
with $\chi^2_{\rm points} = 2.75, 1.55, 1.61$ for jet, pion and kaon data subsets.
The shaded areas, computed following the procedure described in the Appendix of Ref.~\cite{Anselmino:2008sga}, correspond to a 2$\sigma$ uncertainty, and result in the statistical errors quoted in Table~\ref{fitpar}.

\begin{figure}[!thb]
\centering
\includegraphics[trim =  150 0 300 60,clip,width=3.cm]{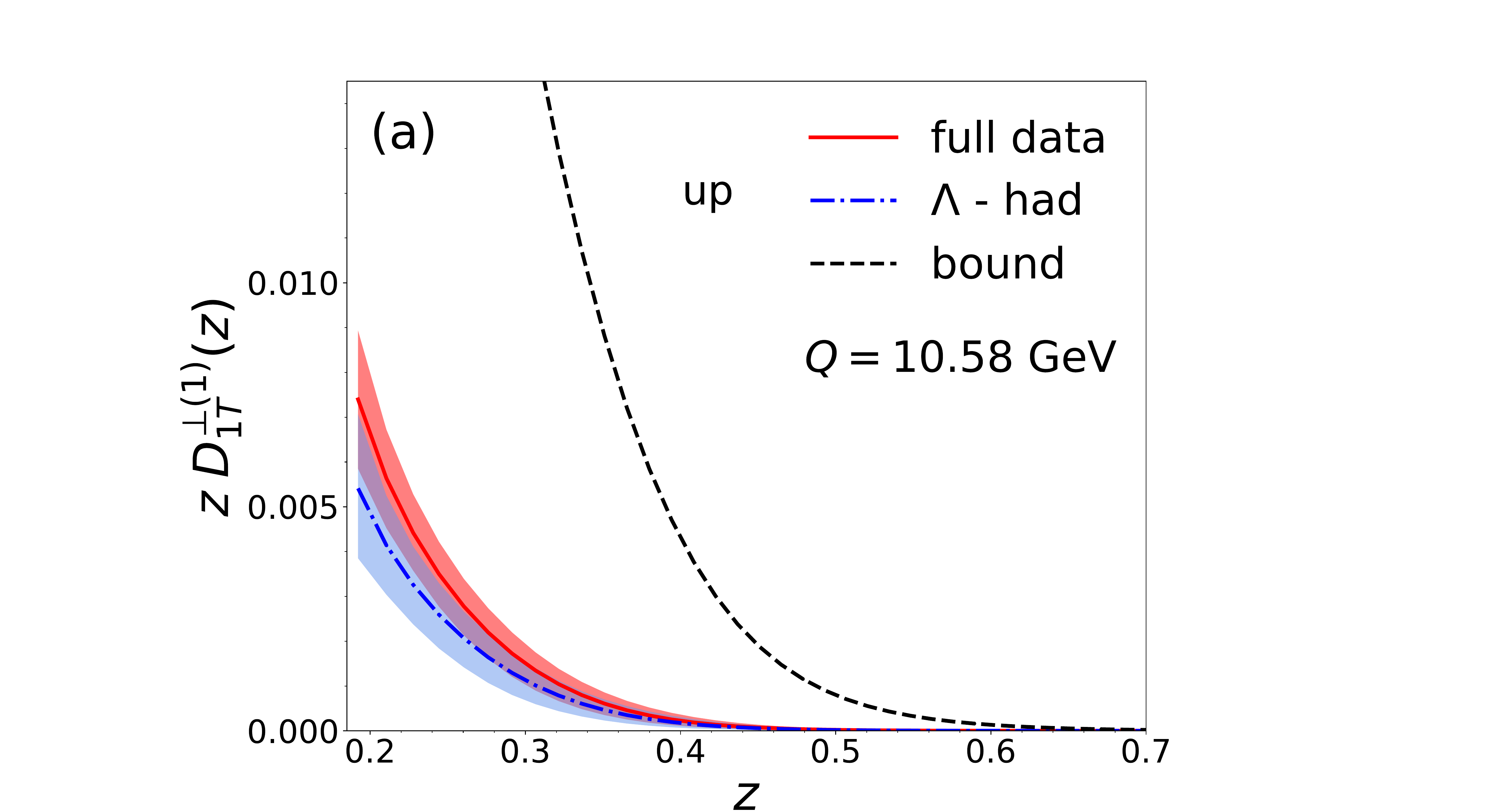}
\includegraphics[trim =  150 0 300 60,clip,width=3.cm]{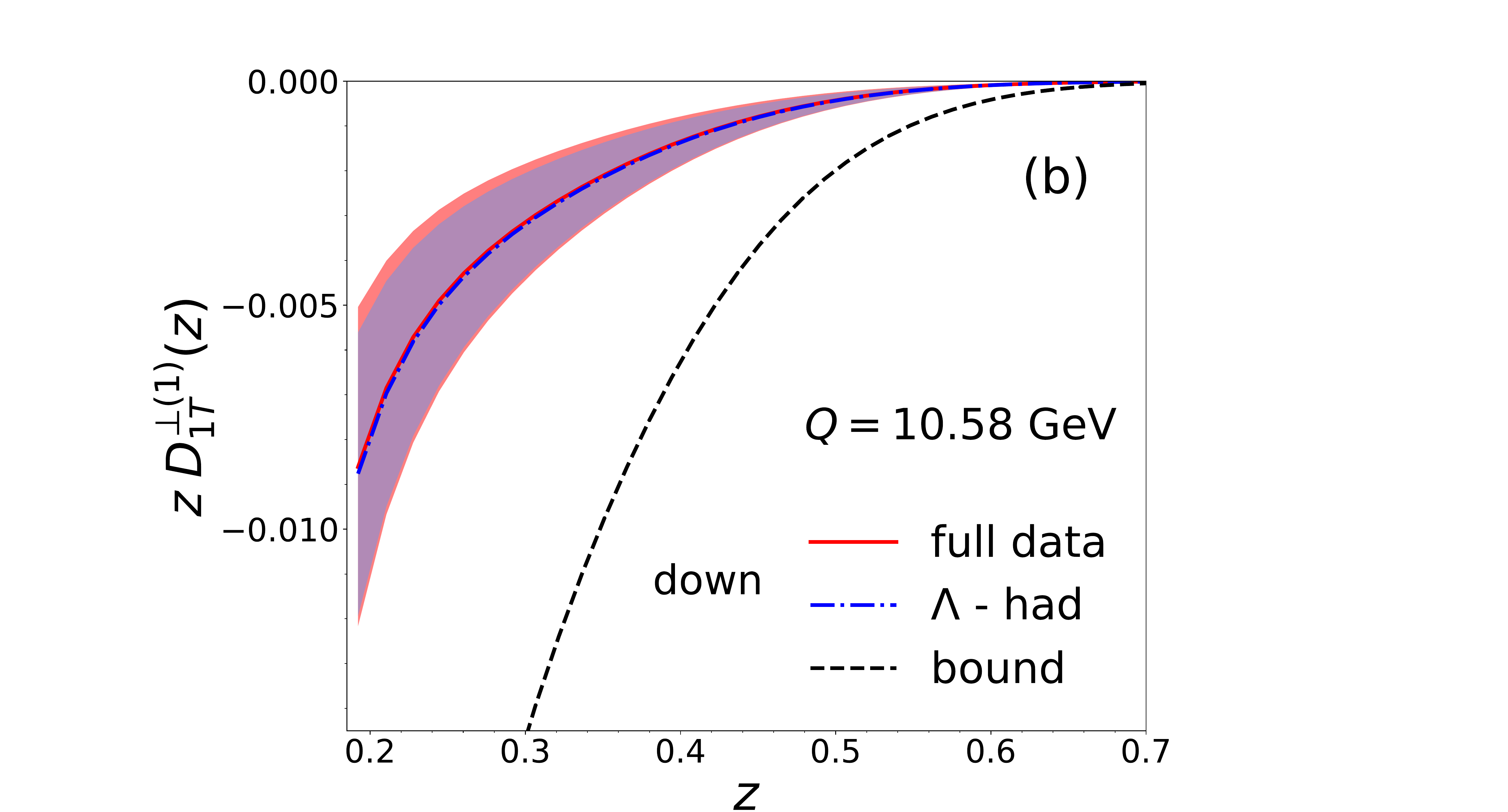}
\includegraphics[trim =  150 0 300 60,clip,width=3.cm]{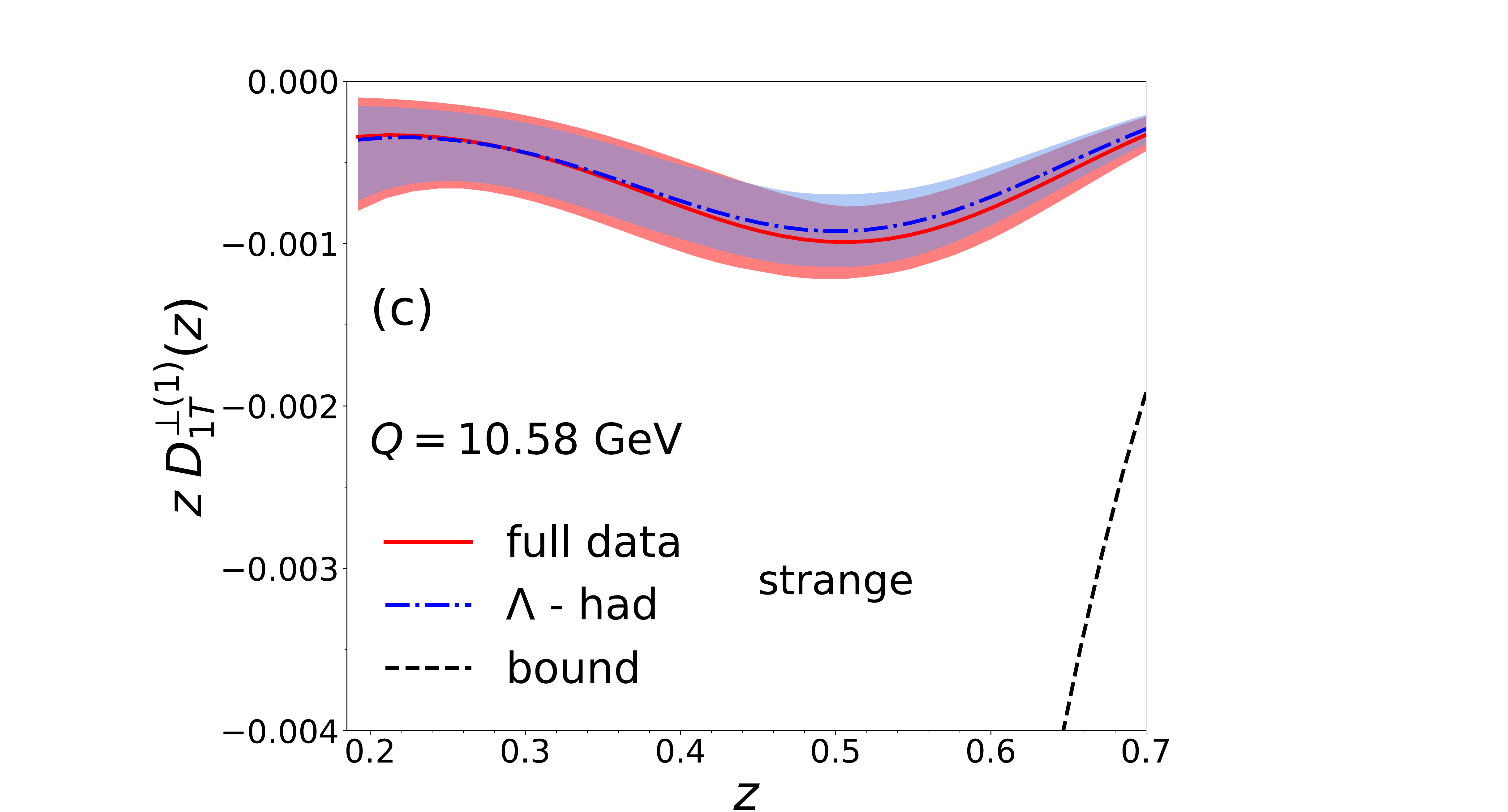}
\includegraphics[trim =  150 0 300 60,clip,width=3.cm]{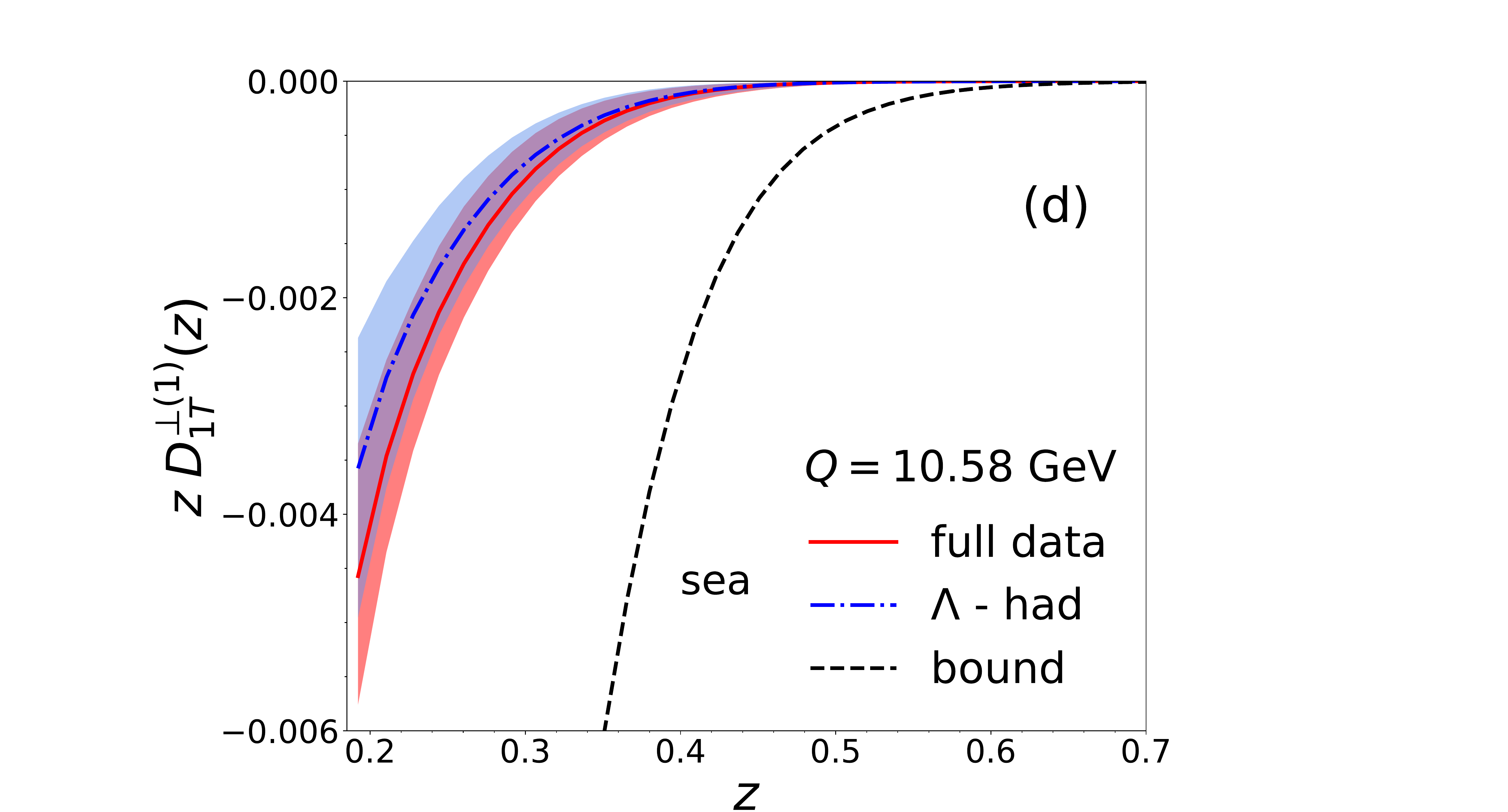}
\caption{First moments of the pFFs, see Eq.~(\ref{1mom}), for the up (a), down (b), strange (c) and sea (d) quarks, as obtained from the full-data fit (red solid lines) and the $\Lambda$-hadron fit (blue dot-dashed lines). The statistical uncertainty bands (at 2$\sigma$ level), as well as the positivity bounds (black dashed lines), are shown.}
\label{fig:1stm}
\end{figure}

Even if the best-fit parameters, as obtained in the associated or the full-data fit, are a bit different, the corresponding first moments, Eq.~(\ref{1mom}), are quite stable, as shown in Fig.~\ref{fig:1stm}.

Some comments are in order:
$i)$ The inclusive production case presents the largest $\chi^2$, on the other hand, while one would expect ${\cal P}_T=0$ at $p_\perp=0$, as well as ${\cal P}_T(\bar\Lambda)={\cal P}_T(\Lambda)$, the data do not show such features (Fig.~\ref{fig:Lj});
$ii)$ For the associated production data, the charge-conjugation symmetry, ${\cal P}_n(\Lambda h^+)= {\cal P}_n(\bar\Lambda h^-)$ is clearly visible (Fig.~\ref{fig:Lh});
%
iii) At medium $z_{\pi ,K}$ values, dominated by the valence unpolarized FFs, $\Lambda \pi^-$, $\Lambda \pi^+$ and $\Lambda K^+$ data give direct information on the pFFs respectively for $u$, $d$ and $s$ quarks: ${\cal P}_n(\Lambda \pi^-)>0$ implies a positive pFF of the up quark, while ${\cal P}_n(\Lambda \pi^+)<0$ leads to a negative down quark pFF  (see Figs.~\ref{fig:Lh}a, \ref{fig:1stm}a,b);
$iv)$ For small $z_\Lambda$, sea quark FFs start playing some role, becoming important around $z_\Lambda \le 0.3$;
%
$v)$ The negative values of ${\cal P}_n(\Lambda K^+$) at medium $z_\Lambda$,is driven by a sizeable and negative $\Delta D_{\Lambda^\uparrow\!/s}$ (Fig.~\ref{fig:1stm}c), coupled to the leading FF $D_{K^+/\!\bar s}$. 
Similar reasonings apply to the $\bar\Lambda h$ data set.

\begin{figure}[!thb]
\centering
\includegraphics[width=6.cm]{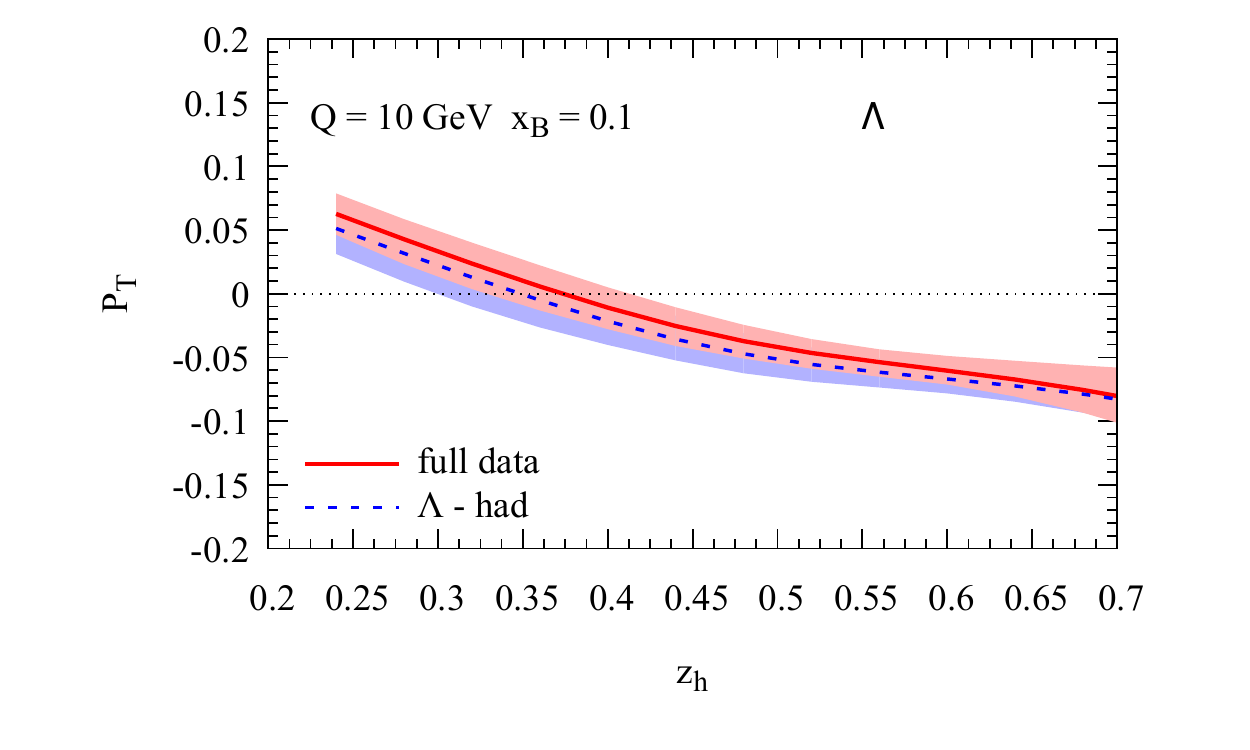}
\includegraphics[width=6.cm]{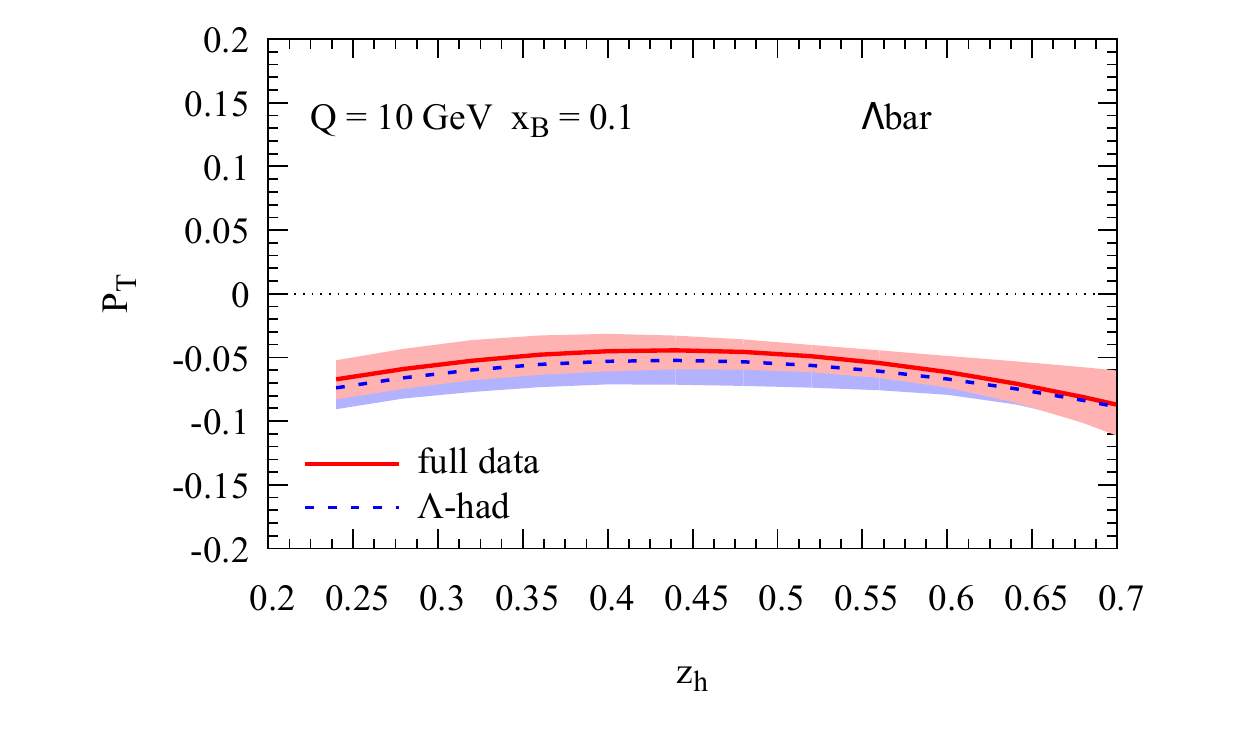}
\caption{Estimates for the transverse polarization of $\Lambda$/$\bar\Lambda$ in SIDIS for EIC kinematics (see legend). The statistical uncertainty bands (at 2$\sigma$ level) are shown.}
\label{fig:sidis}
\end{figure}

The so extracted pFFs can be used to give some predictions for the same observable in SIDIS, in particular for the EIC kinematics. In such a case the polarization is measured transversely w.r.t.~the plane containing the target and the $\Lambda$ particle. We give here directly the final result as a function of $x_{\rm B}$ and $z_h$ (the standard SIDIS variables) and adopting a Gaussian parametrization also for the unpolarized TMD parton distribution:
\be
P(x_{\rm B},z_h) = \frac{\sqrt{2e\pi}}{2m_p}
\frac{\langle p_\perp^2\rangle^2_{\rm pol}}{\langle p_\perp^2\rangle}
\frac{1}{\sqrt {\langle p_\perp^2\rangle_{\rm pol} + z_h^2 \langle k_\perp^2\rangle }} \frac{\sum_q e_q^2 f_{q/p}(x_{\rm B})\Delta D_{\Lambda^\uparrow/q}(z_h)}{\sum_q e_q^2 f_{q/p}(x_{\rm B}) D_{\Lambda/q}(z_h)}\,.
\ee
The corresponding estimates are shown in Fig.~\ref{fig:sidis}, where one can see that the values are sizeable and could allow for a test of the universality of the pFF as well as of its flavor separation.

\section{Conclusions}
\label{concl}

The recent data from Belle Collaboration for the transverse $\Lambda/\bar\Lambda$ polarization have allowed to extract, for the first time,  the polarizing fragmentation function of $\Lambda$ hyperons. Data favor a clear separation in flavors, requiring three different valence pFFs, with their relative sign and size determined quite accurately, as well as a sea-quark pFF.



\paragraph{Funding information}
This work is supported by the European Union's Horizon 2020 research and innovation programme under grant agreement No.~824093 (STRONG2020).



\bibliography{references.bib}

\nolinenumbers

\end{document}